\documentclass[prc,showpacs,floatfix]{revtex4}
\usepackage{bm}
\usepackage{graphicx}
\usepackage{amsmath}

\newcommand{\beq}{\begin{equation}}
\newcommand{\eeq}{\end{equation}}
\newcommand{\beqa}{\begin{eqnarray}}
\newcommand{\eeqa}{\end{eqnarray}}
\begin{document}

\title{
\hfill{\small {\bf MKPH-T-09-07}}\\
{\bf Contribution of single pion electroproduction to the generalized
  Gerasimov-Drell-Hearn sum rule for the deuteron}
}
\author{Alexander Fix$^{a,b}$, Hartmuth Arenh\"ovel$^a$, and Mahmoud Tammam$^c$}
\affiliation{$^a$Institut f\"{u}r Kernphysik, Johannes
Gutenberg-Universit\"{a}t, D-55099 Mainz, Germany,\\
$^b$Laboratory of Mathematical Physics, Tomsk Polytechnic University,
634050 Tomsk, Russia,\\
$^c$Physics Department, Al-Azhar University, Asiut, Egypt
}

\date{\today}

\begin{abstract}
The generalized Gerasimov-Drell-Hearn sum rule is evaluated for the
contribution of single pion production on the deuteron by explicit
integration up to an energy of 1.5~GeV for both coherent and incoherent
production. As elementary $\gamma^* N\rightarrow\pi N$ amplitude the
MAID2003 model has been used. For incoherent production final state
interaction is included in the final $NN$ and $\pi N$ subsystems. The
resulting contribution to the generalized transverse GDH sum rule is
considerably smaller than the negative contribution from the disintegration
channel $d(e,e')np$ which dominates the sum rule at low $Q^2$.
\end{abstract}


\maketitle

\section{Introduction}
\label{sec1}

Because of difficulties in the microscopic description of strongly interacting particles
within quantum chromodynamics, particular significance is attached to relations between
observables obtained directly from general principles of quantum field theory, such as
Lorentz invariance, unitarity, causality {\it etc.} These relations, connecting different
physical aspects of a quantum system, are primarily of strong theoretical significance.
Furthermore, they may be considered as a tool allowing one to check the quality of
different microscopic models, in which the mentioned principles should be taken into
account properly. One of such relations is the Gerasimov-Drell-Hearn
(GDH) sum rule connecting the anomalous magnetic moment of a particle
to the energy weighted integral over the beam-target spin asymmetry of its total
photoabsorption cross section~\cite{Ger65,DrH66}. This sum rule has
become of special interest within the
last 10-15 years because of the substantial progress in the development of polarized
beams and targets. Therefore, considerable success was achieved in measuring the spin
asymmetry for the proton as well as for composite light nuclear systems over a large
energy range, and even data for particular photoabsorption channels
were reported \cite{OJahn}.
Besides the GDH sum rule for real photons, the generalized GDH sum rule for electron
scattering has come into focus in recent years~\cite{Dre01,DrT04}.

Although the existing analyses allow one to conclude that the GDH sum
rule at $Q^2=0$ for nucleons can not be
violated significantly, there are still points needed to be understood. The first one is
the so called neutron puzzle. Whereas the validity of the GDH sum rule for the proton
seems to be confirmed both theoretically and experimentally (at least within the
experimental errors), the situation of the neutron is much less clear. Indeed, an
evaluation of $I_n^{GDH}$ in~\cite{DrT04} (see also calculations of Ref.~\cite{AreFiSch}
and \cite{FiA2pi}) points to a systematic deviation of about 20~\% between the
theoretical evaluation of the integral $I_n^{GDH}$ and its sum rule value. This fact is
quite surprising, especially in view of the much better agreement for the proton within
the same model.

The second point concerns a strong modification of the nucleon spin structure as
one goes from low and moderate $Q^2$-regions, where the generalized GDH sum rule for
the proton $I_p^{GDH}(Q^2)$ is predicted to be large and positive, to the deep inelastic
scattering region, where the value of the integral becomes negative and scales as
$I_p^{GDH}(Q^2)\sim -0.14/Q^2$~\cite{Anselm}. There are general indications that it is
the resonance region which is mainly responsible for this sharp change of the GDH
integral~\cite{Burkert:1992yk}. In this connection it is interesting
to see whether the same strong dependence 
will be exhibited also by the neutron. Thus, the mentioned issues
require an investigation of the generalized GDH sum rule for the
neutron, so that, in view of the absence of free neutron targets,
lightest nuclei, primarily deuteron and $^3$He, become of great
relevance. With respect to nuclear targets, it is important to take into
account disturbing effects from binding and final state interaction (FSI).

The GDH sum rule for the deuteron at $Q^2=0$ has been evaluated in~\cite{AreFiSch} for
all important channels, including photodisintegration, single and double pion
photoproduction as well as eta photoproduction. These channels were found to nearly
saturate the sum rule. The contribution of higher multiple meson final states, being
quite important for the unpolarized cross section, is not expected to be very significant
in the sum rule. This is primarily because of a slowly increasing phase space
available for reactions with more then three particles in the final state. As a result,
multiple pion production starts to come into play at rather high energies $\omega^{lab}
>1$~GeV so that the corresponding contribution to the GDH integral is suppressed by the
weight $1/\omega^{lab}$. The second and probably more important reason
is that with increasing multiplicity of the
final reaction channels, the spin dependence of its amplitude is expected to become less
and less pronounced, so that different spin configurations tend to appear with equal
probability and thus are largely canceled in the sum rule.

With respect to the generalized sum rule of the deuteron, the
contribution of the electrodisintegration channel $d(e,e')np$ has been
evaluated in Ref.~\cite{Are04}. The calculation was based on
a conventional nonrelativistic framework using a realistic $NN$-potential and including
contributions from meson exchange currents, isobar configurations and leading order
relativistic terms. By integrating up to a maximal internal excitation energy of the
final $np$-system $E_{np}=1$~GeV, good convergence was achieved. The prominent feature of
the electrodisintegration channel to the generalized GDH sum rule as function of the
squared four-momentum $Q^2$ is a pronounced deep negative minimum, $I_{\gamma^* d\to
np}^{GDH}=-9.5$~mb, at low $Q^2\approx 0.006$~(GeV/c)$^{2}$ (see
Fig.~\ref{fig_gen_gdh_int_v18_piall}) which is almost exclusively driven by the nucleon
isovector anomalous magnetic moment contribution to the magnetic dipole transition to the
$^1S_0$ scattering state. Above $Q^2=0.8$~(GeV/c)$^{2}$ the integral
$I^{GDH}_{\gamma^*d\to np}(Q^2)$ approaches zero rapidly.

Apart from the aspects related to the neutron, the generalized GDH sum rule for the
deuteron is interesting by itself. Indeed, as has been shown in in Ref.~\cite{AreFiSch},
an almost vanishing value for the GDH integral at $Q^2=0$ according to
the sum rule value $I_d^{GDH}=0.65~\mu$b as dictated by the very small
anomalous magnetic moment of the deuteron is provided by an almost exact cancelation
between nucleon degrees of freedom (photodisintegration channel) and
subnucleon degrees of freedom (mainly pion as manifest in single and double pion
photoproduction). In view of the large
negative contribution from the photodisintegration channel of
about -380~$\mu$b, the mentioned cancelation has to eliminate the three
leading decimals. This feature requires a consistent treatment of
nucleon, pion and other subnucleon degrees of freedom.

In view of the above mentioned dramatic change with increasing $Q^2$
of the contribution $I_{\gamma^* d\to np}^{GDH}(Q^2)$ from
electrodisintegration to the generalized GDH sum rule,
going through a deep minimum, the
natural question arises, whether this large negative contribution will
be canceled again by the contribution of
meson electroproduction similar to the case of real
photons. Clearly, to that end the spin asymmetry of electroproduction on the deuteron
must strongly increase above the pion
production threshold. On the one hand, the strength of magnetic transitions, dominating
pion production at lower energies increases with $Q^2$, at least as
long as a suppression
of the long range mechanisms (resonance excitation) does not start to come into play in
the region of high momentum transfers. On the other hand, with increasing $Q^2$ the nuclear
structure becomes more effective in leading to a decrease of the reaction rate via the
nuclear form factor, which must be particularly pronounced in the case
of a deuteron, having a rather extended structure. Thus the question
about the behavior of $I_d^{GDH}(Q^2)$ at
$Q^2>0$ is by no means trivial and must be studied within a sufficiently refined
model. The latter has to take into account correctly nuclear effects, such as Fermi
motion, off-mass shell corrections and final state interactions.

The aim of the present paper is an evaluation of the leading contribution above pion
production threshold to the generalized GDH integral $I_d^{GDH}(Q^2)$,
namely incoherent and coherent single pion electroproduction, $d(e,e'\pi)NN$ and
$d(e,e'\pi^0)d$. These reactions were considered in detail in Ref.~\cite{TaF06}. Special
attention had been given to polarization observables and to the role of $NN$ and
$\pi N$ interactions in the final state. The calculation requires proper treatment of the
elementary electroproduction reaction $N(e,e^\prime\pi)N$. The physical picture
underlying the electroproduction in the region of low and medium energies is usually
presented in terms of transitions from the nucleon to $N^*$ and
$\Delta$ resonances. These have nonperturbative character and,
therefore, need a phenomenological
model for their description. In Ref.~\cite{TaF06} the MAID2003 analysis
\cite{MAID03} was used, which describes electroproduction of pions via excitation of
$s$-channel resonances with nonresonant contributions from the nucleon poles as well as
meson exchange in the $t$-channel. Utilizing MAID2003 developed up to a total c.m.\
energy $W=2$~GeV, the calculation on the deuteron in~\cite{TaF06} could be extended up to
a photon lab energy $\omega^{lab}=1.5$~GeV, thus covering the major part of the resonance
region.

As for $NN$ and $\pi N$ FSI effects, it turned out that their influence is similar
to that noted previously for the photoproduction processes in Ref.~\cite{AreFiSch}.
Namely, the most significant role is played by the $np$ interaction in the $^3S_1$ state
in the neutral channel $d(e,e'\pi^0)np$, resulting in a visible reduction of the
corresponding impulse approximation (IA). In the charged channels, $NN$ FSI leads to much
smaller effects, about 1-2~\% in the first resonance region. The origin of such a
different role of FSI in the charged and neutral channels was found in a spurious
contribution of the coherent process to the incoherent process in IA. This spurious
contribution is admixed unavoidably to the incoherent one if the $np$
FSI is neglected, because of the nonorthogonality between the plane
wave of the impulse approximation and the bound state wave
function. After elimination of this spurious contribution by a
projection technique, the
remaining FSI effect is comparable in size to that seen in the
$\pi^+nn$ and $\pi^-pp$ channels. The $\pi N$ rescattering is insignificant over the
whole energy region $\omega^{lab}\leq 1.5$ GeV. It is also worth noting that although in
the unpolarized cross section FSI may safely be neglected at least not very close to the
threshold, their role may appear to be important in polarization observables, especially
in the target polarization, as was shown in Ref.~\cite{TaF06}.

In the next two Sections~\ref{sec3} and \ref{sec3a} we
give relevant formulas for the generalized GDH sum rule for the deuteron. In
Sect.~\ref{sec4a} the contribution to the sum rule from single pion electroproduction is
presented. In Sect.~\ref{sec4} we summarize the results and present some conclusions.

\section{The generalized GDH sum rule}
\label{sec3}

\begin{figure}[htb]
\includegraphics[scale=.6]{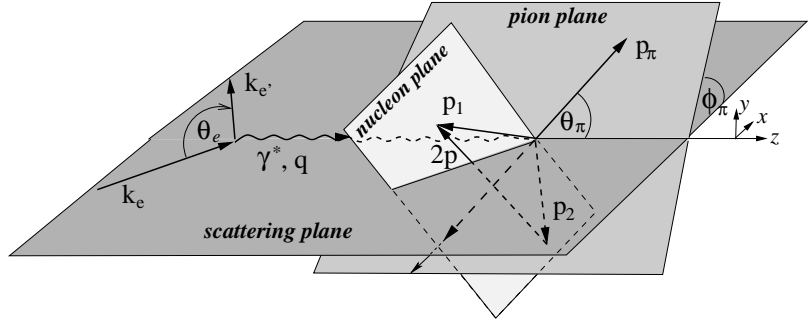}
\caption{Kinematics of single pion electroproduction on the deuteron
  in the $\gamma^*d$ cm.\ system.}
\label{fig_elpion_kinematik}
\end{figure}

The generalized GDH sum rule is determined by the vector beam-target asymmetry $A_{ed}^V$
of the inclusive electron deuteron scattering cross section for longitudinally polarized
electrons. For electroproduction the latter has the form~\cite{TaF06} (correcting some
misprints)
\beqa
\sigma_{e,\pi}(h,P^d_1,P^d_2,\theta_d,\phi_d)&\equiv& \frac{d^3\sigma}{dE_{e'}
d\Omega_{e'}}\nonumber\\&=& \frac{\alpha_{qed}}{Q^4}\,\frac{k_{e'}}{k_e}
\Big[\rho_{L} F_{L}^{00}+\rho_T F_{T}^{00} 
+P^d_1(h\rho_T' F_{T}^{\prime 10}\cos\theta_d+
[h\rho_{LT}' F_{LT}^{\prime 11}\cos\phi_d-\rho_{LT} F_{LT}^{11}\sin\phi_d]
d^1_{10}(\theta_d)) \nonumber\\
&&+P^d_2([\rho_{L} F_{L}^{20}+\rho_T F_{T}^{20}]d^2_{00}(\theta_d) +[\rho_{LT}
F_{LT}^{21}\cos\phi_d-\rho_{LT}' F_{LT}^{\prime 21}
\sin\phi_d]d^2_{10}(\theta_d)\nonumber\\
&& +\rho_{TT} F_{TT}^{22}d^2_{20}(\theta_d)\cos (2\phi_d))\Big]\,, \label{inclusive_xs}
\eeqa
with $\alpha_{qed}$ as electromagnetic fine structure constant
and where incoming and scattered
electron energies and momenta are denoted by $E_e$, $k_e$ and $E_{e'}$, $k_{e'}$,
respectively. Furthermore, $\rho_\alpha^{(\prime)}$ with $\alpha \in\{L,T,LT,TT\}$ denote
the virtual photon polarization parameters, $P^d_1$ and $P^d_2$ the vector and tensor
deuteron polarization parameters, respectively, $(\theta_d,\phi_d)$
the spherical angles of the
deuteron orientation axis, and $h$ the degree of longitudinal electron
polarization. The kinematics is displayed in Fig.~\ref{fig_elpion_kinematik}.

The various form factors $F_\alpha^{(\prime)IM}$ are defined in~\cite{TaF06} and the
reader is referred to this work for more details. They are functions of the energy transfer
$\omega$ in the $\gamma^*d$ c.m.-system and the squared four momentum transfer
$Q^2=-q_\mu^2$. At the photon point, the transverse form factors are related to the
corresponding asymmetries of the total photoproduction cross section as listed in
eq.~(83) of~\cite{ArF05}
\beq F_T^{00}=\frac{\omega\,W}{\pi^2 E_d}\sigma^0\,,\quad
\frac{F_{T}^{20}}{F_T^{00}}=\overline T_{20}^{\,0}\,,\quad \frac{F_{T}^{\prime
10}}{F_T^{00}} =\overline T_{10}^{\,c}\,, \quad \frac{F_{TT}^{22}}{F_T^{00}}=\overline
T_{22}^{\,l}\,,\label{relations}
\eeq
where $\sigma^0$ denotes the unpolarized inclusive
photoproduction cross section, $\overline T_{20}^{\,0}$ the tensor target asymmetry,
$\overline T_{10}^{\,c}$ the beam-target vector asymmetry for circularly polarized
photons and $\overline T_{22}^{\,l}$ the beam-target tensor asymmetry for linearly
polarized photons. (N.B.: In the corresponding relations listed in eq.~(60)
of~\cite{TaF06} a factor $\sigma^0$ is missing on the right hand sides of $F_{T}^{20}$,
$F_{T}^{\prime 10}$, and $F_{TT}^{22}$.)

The various beam, target and beam-target asymmetries of inclusive
scattering are defined by the following general form of the inclusive cross section
\beqa
\sigma_{e,\pi}(h,P^d_1,P^d_2,\theta_d,\phi_d)&=&\sigma_{e,\pi}^0
\Big(1+P^d_1\,A_d^V(\theta_d,\phi_d) +P^d_2\,A_d^T(\theta_d,\phi_d)
\nonumber\\ &&
+h[A_{ed}(\theta_d,\phi_d)+P^d_1\,A_{ed}^V(\theta_d,\phi_d)
+P^d_2\,A_{ed}^T(\theta_d,\phi_d)]\Big)\,, \label{asymmetries}
\eeqa
where
\beq
\sigma_{e,\pi}^0=\frac{\alpha_{qed}}{Q^4}\,\frac{k_{e'}}{k_e} \Big(\rho_{L}
F_{L}^{00}+\rho_T F_{T}^{00}\Big)
\eeq
denotes the unpolarized inclusive scattering cross section. By proper
choices of the polarization parameters $h$, $P^d_1$, and $P^d_2$, one
can separate the various asymmetries. For example, the beam-target
vector asymmetry is obtained by combining four different settings of
the polarization parameters according to
\beqa
A_{ed}^V(\theta_d,\phi_d)&=&\frac{1}{4\,h\,P_1^d\sigma_e^0}
{\Big[}(\sigma_e(h,P_1^d,P_2^d,\theta_d,\phi_d)
-\sigma_e(-h,P_1^d,P_2^d,\theta_d,\phi_d)\nonumber\\&&
-\sigma_e(h,-P_1^d,P_2^d,\theta_d,\phi_d)
+\sigma_e(-h,-P_1^d,P_2^d,\theta_d,\phi_d)\Big]\,.
\eeqa

Comparing (\ref{asymmetries}) with (\ref{inclusive_xs}), one can
express the asymmetries in terms of form factors and kinematic quantities.
For the vector asymmetry $A_{ed}^V$ one obtains
\beq
A_{ed}^V(\theta_d,\phi_d)=\frac{\alpha_{qed}}
{Q^4\,\sigma_{e,\pi}^0}\,\frac{k_{e'}}{k_e}
\Big(\rho_T' F_{T}^{\prime 10}\cos\theta_d
+\rho_{LT}' F_{LT}^{\prime 11}\cos\phi_d\,d^1_{10}(\theta_d)\Big)\,,
\eeq
yielding for $(\theta_d,\phi_d)=(0,0)$, i.e.\ deuteron orientation
axis parallel to $\vec q$,
\beq
A_{ed}^V(0,0)=\frac{\alpha_{qed}}
{Q^4\,\sigma_{e,\pi}^0}\,\frac{k_{e'}}{k_e}
\rho'_T F^{\prime 10}_T\,.
\eeq
The expression $2\,\sigma_{e,\pi}^0A_{ed}^V(0,0)$ describes exactly the
total electroproduction cross section asymmetry for completely
polarized electrons ($h=1$) and
complete deuteron spin aligned parallel and antiparallel to the
momentum transfer, which means $P^d_1=\pm\sqrt{3/2}$, respectively, and
$P^d_2=1/\sqrt{2}$. Because defining
\beqa
\sigma_{e,\pi}^{P/A}&=&\sigma_{e,\pi}(1,\pm\sqrt{\frac{3}{2}},\frac{1}{\sqrt{2}},0,0)
\nonumber\\
&=&\frac{\alpha_{qed}}{Q^4}\,\frac{k_{e'}}{k_e}
\Big[\rho_{L} (F_{L}^{00}+\frac{1}{\sqrt{2}}F_{L}^{20})
+\rho_T (F_{T}^{00}+\frac{1}{\sqrt{2}}F_{T}^{20})
\pm\sqrt{\frac{3}{2}}\rho'_T F^{\prime 10}_T\Big]\,,
\eeqa
one finds for the cross section asymmetry
\beq
\sqrt{\frac{2}{3}}(\sigma_{e,\pi}^{P}-\sigma_{e,\pi}^{A})=2\frac{\alpha_{qed}}
{Q^4}\,\frac{k_{e'}}{k_e}\rho'_T F^{\prime 10}_T\,.
\eeq
In view of the spin asymmetry for real photons
\beq
\sigma_\gamma ^P(\omega^{lab})-\sigma_\gamma ^A(\omega^{lab})
=2\sigma_0\,\overline T_{10}^{\,c}(\omega^{lab})=
2\frac{\pi^2\,E_d}{{W\,q}}F^{\prime 10}_T(W,Q^2=0)\,,
\label{spin_as_real}
\eeq
using the expressions in Eq.~(\ref{relations}),
we introduce for transverse virtual photons the parallel and
antiparallel spin aligned cross sections
\beq
\sigma_{T,\gamma^*} ^{P/A}(\omega^{lab},Q^2)=\frac{\pi^2\,E_d}{{W\,q}}
\Big[F_{T}^{00}(\omega^{lab},Q^2)+\frac{1}{\sqrt{2}}F_{T}^{20}(\omega^{lab},Q^2)
\pm\sqrt{\frac{3}{2}}\frac{\rho'_T}{\rho_T} F^{\prime 10}_T(\omega^{lab},Q^2)\Big]\,,
\eeq
where ${\rho'_T}/{\rho_T}$ describes the degree of circularly
polarized virtual photons. Accordingly, we introduce
as spin asymmetry for transverse virtual photons
\beq
\Sigma_{\gamma^*}(\omega^{lab},Q^2)=\sqrt{\frac{2}{3}}\frac{\rho_T}{\rho'_T}
(\sigma_{T,\gamma^*} ^P(\omega^{lab},Q^2)
-\sigma_{T,\gamma^*} ^A(\omega^{lab},Q^2))=
2\frac{\pi^2\,E_d}{{W\,q}}F^{\prime 10}_T(W,Q^2)
\,,\label{spin_asy_virtual}
\eeq
which coincides at the photon point with (\ref{spin_as_real}).
Correspondingly, we take as extension of the GDH integral from real
to virtual photons the definition
\beqa
I_d^{GDH}(Q^2)&=&\int_{\omega_{th}^{lab}}^\infty
\frac{d\omega^{lab}}{\omega^{lab}}\,
\Sigma_{\gamma^*}(\omega^{lab},Q^2)\nonumber\\
&=&2\,\pi^2\,\int_{\omega_{th}^{lab}}^\infty
\frac{d\omega^{lab}}{\omega^{lab}}\,\frac{E_d}{W\,q}
F_{T} ^{\prime 10}(W,Q^2)\,g(\omega^{lab},Q^2)\,,
\label{gdh_virtual}
\eeqa
where $W$ denotes the invariant mass which is a function of $\omega$
and $Q^2$
\beq
W=\omega+\sqrt{M_d^2+q^2}=\sqrt{(2\omega^{lab}+M_d)M_d-Q^2}
\eeq with $M_d$ as deuteron mass. The threshold invariant mass is given by
$W_{th}=2M+m_\pi$ with $M$ and $m_\pi$ as nucleon and pion masses, respectively, and thus
the threshold lab energy by $\omega_{th}^{lab}=(W_{th}^2+Q^2-M_d^2)/2M_d$. Furthermore,
$E_d$ and $(\omega,\vec q\,)$ denote the deuteron energy and the virtual photon
four-momentum in the $\gamma^*d$ c.m. system, respectively. The factor
$g(\omega^{lab},Q^2)$ in (\ref{gdh_virtual}) takes into account the fact, that the
generalization of the GDH integral is to a certain extent arbitrary. The only restriction
for this factor is the condition that at the photon point $Q^2=0$ one
has
\beq
g(\omega^{lab},0)=1\,,
\eeq
and that
\beq \lim_{\omega^{lab}\rightarrow
\infty}g(\omega^{lab},Q^2)|_{Q^2=const.}<\infty
\eeq
remains finite. As simplest
extension we choose here $g(\omega^{lab},Q^2)\equiv 1$. For the
explicit integration over a finite range of $\omega^{lab}$ we introduce the finite
GDH integral by
\beqa
I_d^{GDH}(Q^2,\omega^{lab}_{max})
&=&2\,\pi^2\int_{\omega_{th}^{lab}}^{\omega^{lab}_{max}}
\frac{d\omega}{\omega}\,\frac{E_d}{W\,q} F_{T} ^{\prime 10}(W,Q^2)\,.
\label{gdh_virtual_finite}
\eeqa

Transforming (\ref{gdh_virtual}) into an integral over $W$, using
\beq
\omega^{lab}=\frac{1}{2M_d}\,(W^2+Q^2-M_d^2)\,,
\eeq
one obtains
\beq
I_d^{GDH}(Q^2)=4\pi^2\int_{W_{th}}^\infty dW
\frac{E_d(W,Q^2)}{q(W,Q^2)}\,\frac{F_{T}^{\prime 10}(W,Q^2)}{(W^2+Q^2-M_d^2)}\,,
\label{gdh_virtual_a}
\eeq
where now $E_d$ and $q$, the three-momentum in the $\gamma^*d$
c.m.-system, have to be considered as functions of $W$ and $Q^2$, i.e.\
\beqa
E_d(W,Q^2)&=&\sqrt{M_d^2+q^2(W,Q^2)}\,,\\
q(W,Q^2)&=&\frac{1}{2W}\,\sqrt{((W-M_d)^2+Q^2)
((W+M_d)^2+Q^2)}\,.
\eeqa

\section{The Reaction Amplitude}
\label{sec3a}

In the present work the contribution of single pion production to the
generalized GDH sum rule of (\ref{gdh_virtual_a}) is evaluated by explicit
integration up to a maximal lab virtual photon energy
$\omega^{lab}_{max}=1.5$~GeV. The evaluation is based on the formalism
developed in Ref.~\cite{TaF06} which we will briefly review.
\begin{figure}[tbph]
\centerline{\includegraphics[width=.49\textwidth]{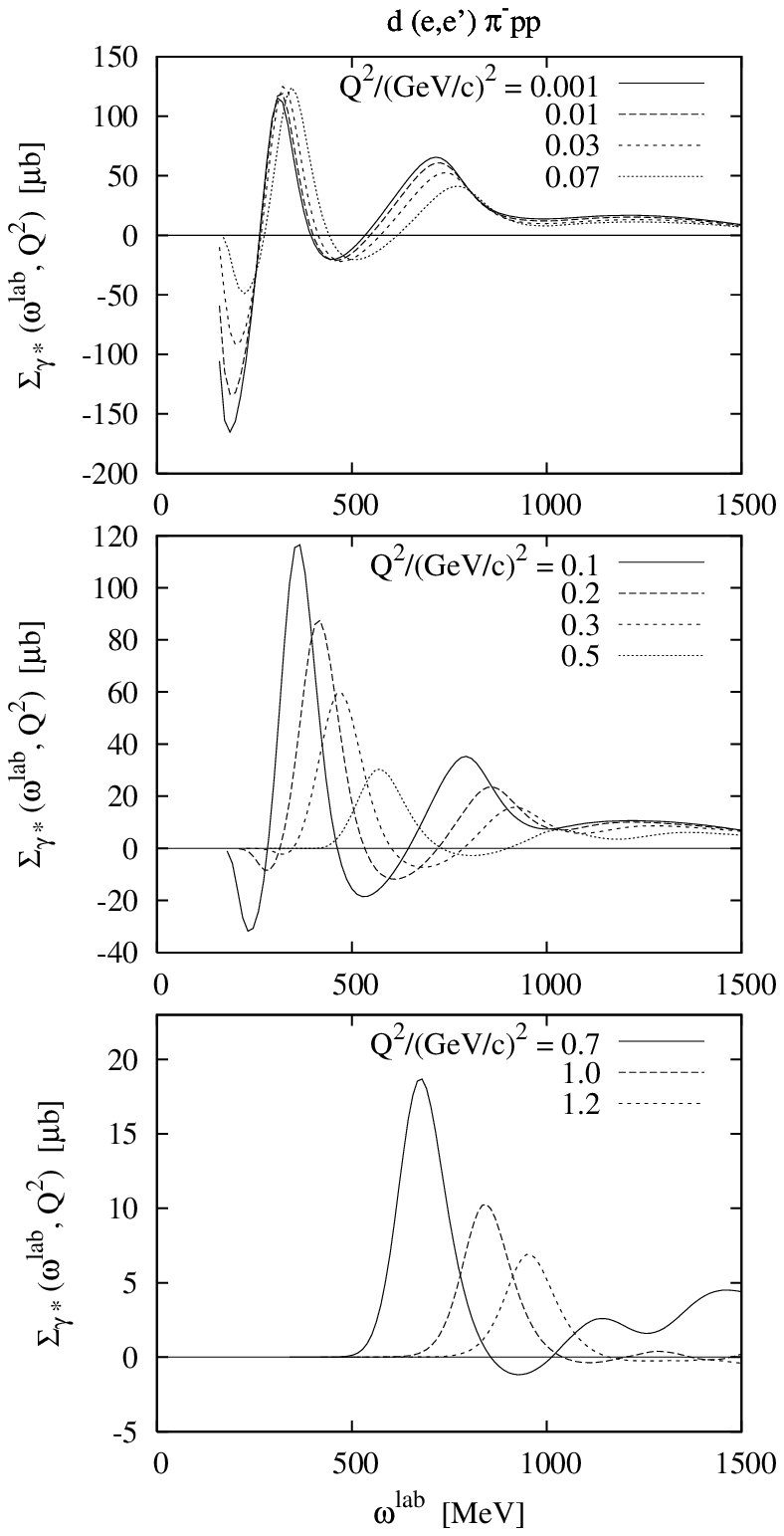}
\includegraphics[width=.49\textwidth]{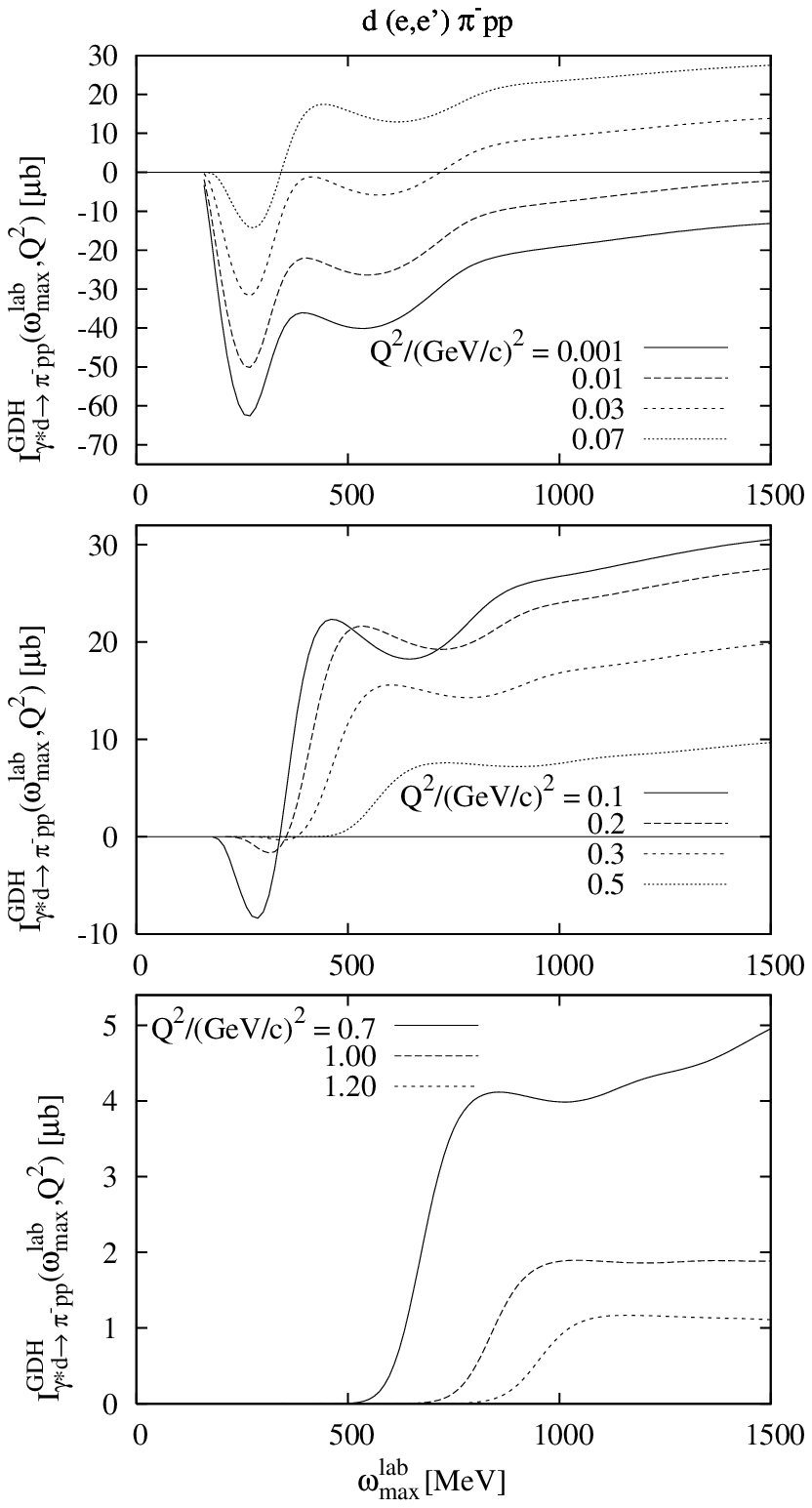}}
\caption{Transverse spin asymmetry $\Sigma_{\gamma^*d\to\pi^-pp}$  as
  function of $\omega^{{lab}}$ (left panel)
  and finite GDH integral $I^{GDH}_{\gamma^*d\to\pi^-pp}$ as function of
  $\omega^{{lab}}_{max}$ (right panel) of $\pi^-$ electroproduction on the deuteron
$d(e,e'\pi^-)pp$ for various constant squared
four-momentum transfers $Q^2$.} \label{fig_gen_gdh_pim}
\end{figure}
The general form of the $T$-matrix is given by
\beqa
T_{s m_s \mu m_d}(W,Q^2,p_\pi,\Omega_\pi,\Omega_p)&=& -
^{(-)}\langle \vec p\, s m_s,\,\vec p_\pi\,|J_{\gamma\pi,\,\mu}(\vec
q\,)|1m_d\rangle\nonumber\\
&=& \sqrt{2\pi}\sum_{L} i^L\hat L ^{(-)}\langle \vec p\, s m_s,\,\vec p_\pi\, |{\cal
O}^{\mu L}_\mu|1m_d\rangle\,,
\eeqa
where $s$ and $m_s$ denote the total spin and its
projection on the relative momentum $\vec p$ of the outgoing two nucleons, and $m_d$
correspondingly the deuteron spin projection on the $z$-axis as quantization axis.
Furthermore, $\mu\in\{0,\pm 1\}$ enumerates the spherical current components with the
provision that $J_{\gamma\pi,\,0}$ is identified with the charge density. We use
through out the notation $\hat L=\sqrt{2L+1}$. The kinematic quantities and the geometry
is explained in Fig.~\ref{fig_elpion_kinematik}. The symbol ${\cal O}^{\mu L}_M$ denotes
charge ($C_M^L$) and transverse multipoles ($E_M^L$ and $M_M^L$)
according to
\begin{eqnarray}
{\cal O}^{\mu L}_M&=& \delta_{\mu 0}C_M^L +\delta_{|\mu| 1}(E_M^L +\mu
M_M^L)\,.
\end{eqnarray}
Using a partial wave decomposition of the final states, one can separate the
$\phi_\pi$-dependence
\begin{eqnarray}\label{small_t}
T_{s m_s \mu m_d}(W,Q^2,p_\pi,\Omega_\pi,\Omega_p)&=&
e^{i(\mu+m_d-m_s)\phi_\pi}
t_{s m_s \mu m_d}(W,Q^2,p_\pi,\theta_\pi,\theta_p,\phi_{p\pi})\,,
\end{eqnarray}
where the small $t$-matrix depends besides $W$, $Q^2$ and the pion
momentum $p_\pi$ only
on $\theta_\pi$, $\theta_p$, and the relative azimuthal angle
$\phi_{p\pi}=\phi_p-\phi_\pi$. For further details see Ref.~\cite{TaF06}.

\begin{figure}[htbp]
\centerline{\includegraphics[width=.49\textwidth]{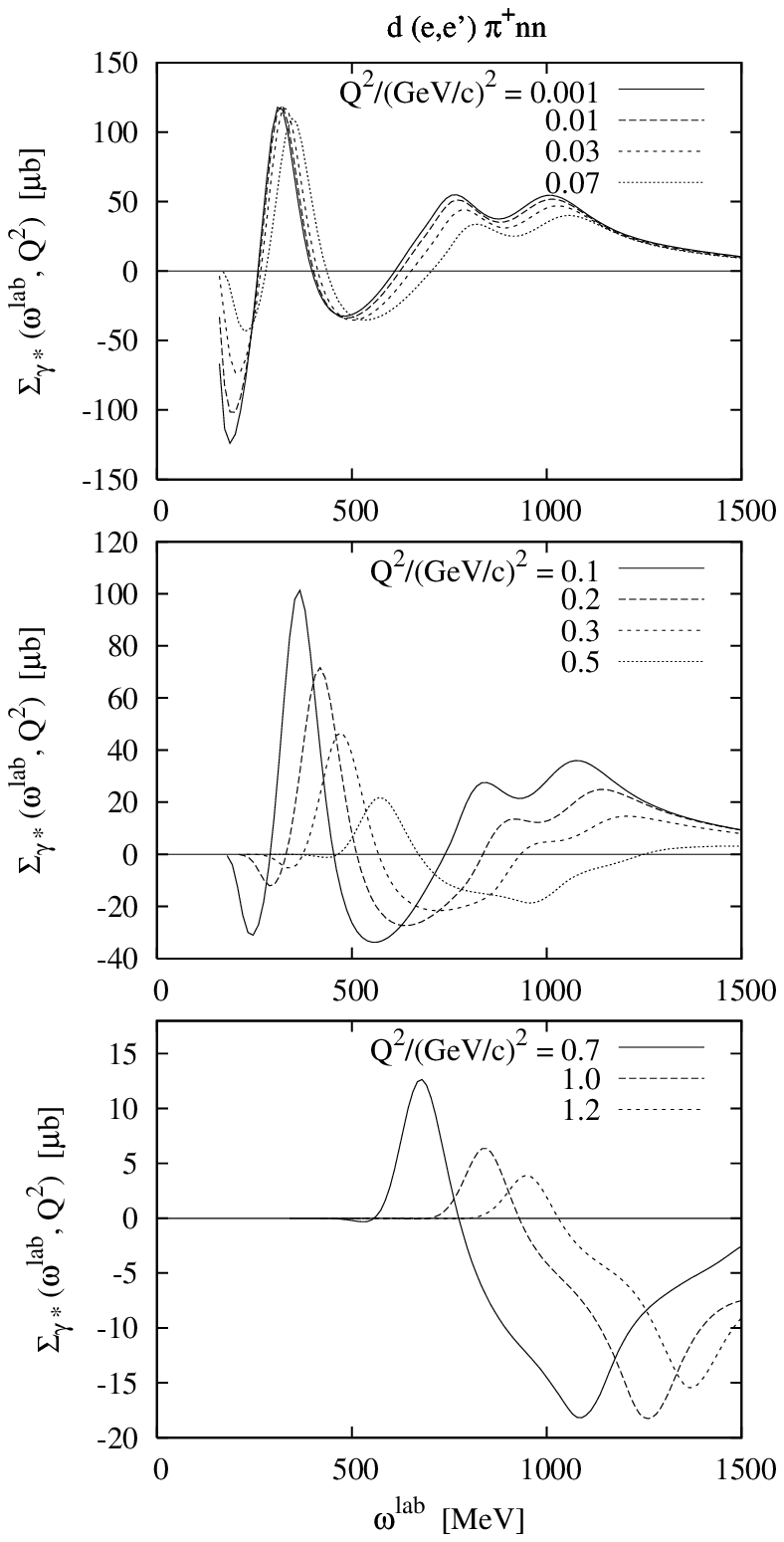}
\includegraphics[width=.49\textwidth]{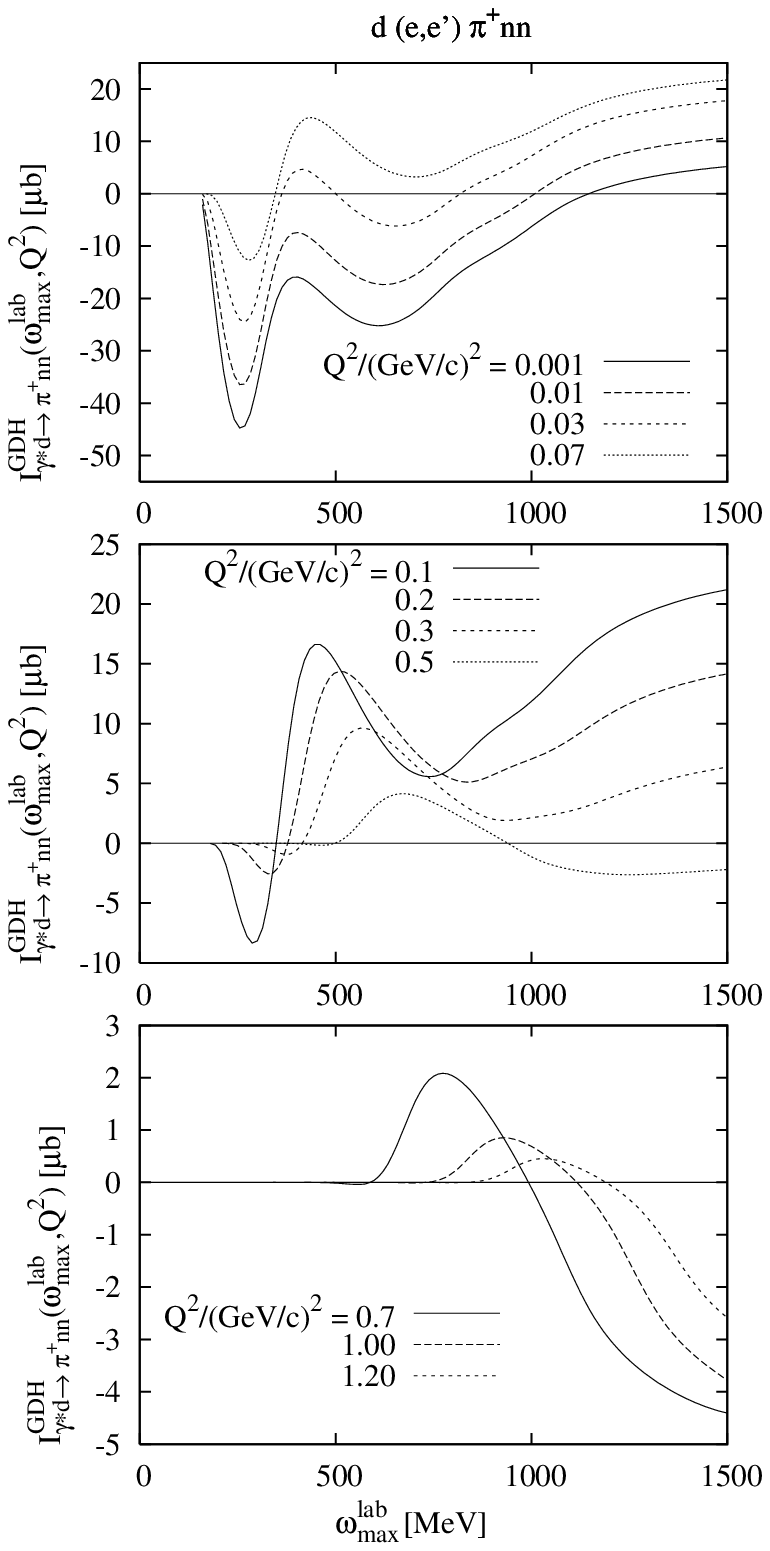}}
\caption{Transverse spin asymmetry $\Sigma_{\gamma^*d\to\pi^+nn}$  as
  function of $\omega^{{lab}}$ (left panel)
  and finite GDH integral $I^{GDH}_{\gamma^*d\to\pi^+nn}$ as function of
  $\omega^{{lab}}_{max}$ (right panel) of $\pi^+$ electroproduction on the deuteron
$d(e,e'\pi^+)nn$ for various constant squared
four-momentum transfers $Q^2$.} \label{fig_gen_gdh_pip}
\end{figure}

According to (\ref{gdh_virtual}) the form factor $F_{T}^{\prime 10}$ is the relevant
quantity which determines the generalized GDH sum rule. It is expressed in terms of the
small $t$-matrix elements
via integration over the final phase space
\begin{equation}
F_{T}^{\prime 10}=\frac{1}{\sqrt{6}}\int dp_\pi d\Omega_\pi
d\Omega_p\,c(W,Q^2,p_\pi,\Theta_\pi,\Theta_p,\phi_{\pi
p})\sum\limits_{sm_s}\Big(|t_{sm_s 1 1}|-|t_{sm_s 1 -1}|\Big)\,,
\end{equation}
where
\begin{equation}
c(W,Q^2,p_\pi,\theta_\pi,\theta_p,\phi_{p\pi})=\frac{M^2p^2p_\pi^2}
{8(2\pi)^4E_\pi(E_{12}p+\frac{1}{2}p_\pi(E_1-E_2)\cos{\theta_{p\pi}})}\,.
\end{equation}
denotes a kinematic phase space factor in which all kinematic quantities refer to the
$\gamma^*d$ c.m.\ system ($E_\pi$ and $p_\pi$ pion energy and
momentum, $E_1$ and $E_2$ nucleon
energies, and $p$ relative nucleon momentum, see Fig.~\ref{fig_elpion_kinematik}).

\begin{figure}[htbp]
\centerline{\includegraphics[width=.49\textwidth]{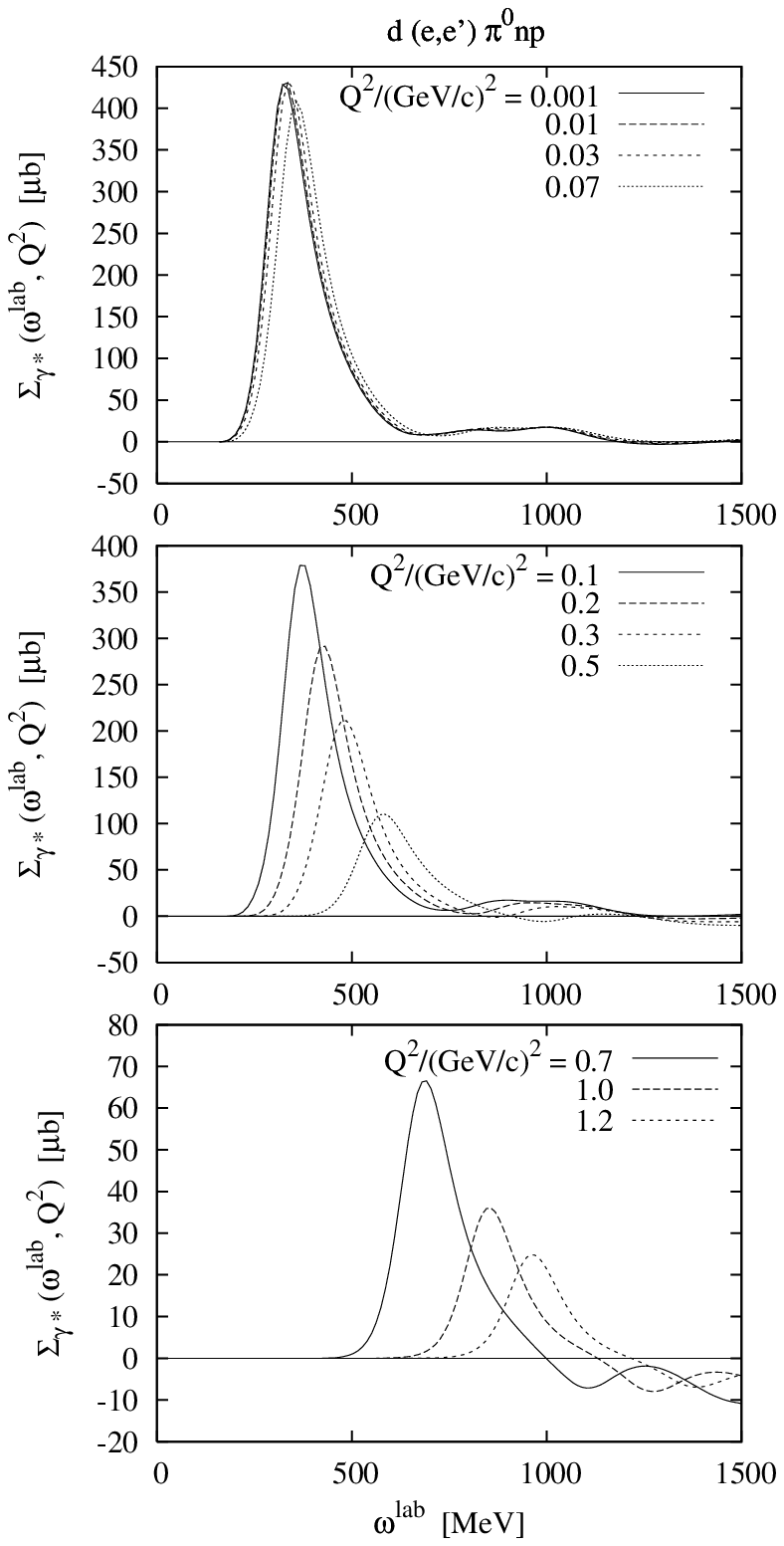}
\includegraphics[width=.49\textwidth]{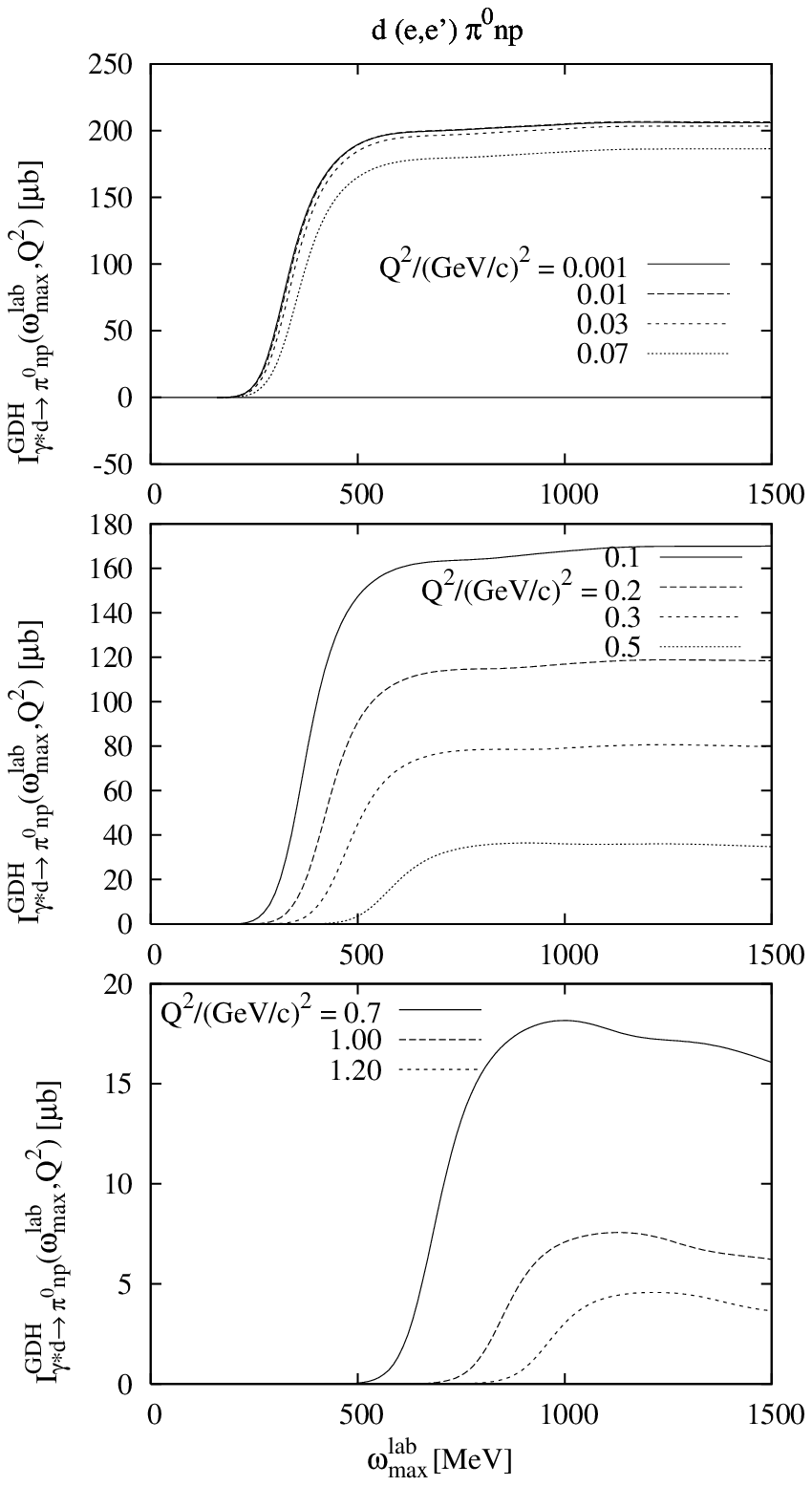}}
\caption{Transverse spin asymmetry $\Sigma_{\gamma^*d\to\pi^0np}$  as
  function of $\omega^{{lab}}$ (left panel)
  and finite GDH integral $I^{GDH}_{\gamma^*d\to\pi^0np}$ as function of
  $\omega^{{lab}}_{max}$ (right panel) of incoherent $\pi^0$
  electroproduction on the deuteron
$d(e,e'\pi^0)np$ for various constant squared
four-momentum transfers $Q^2$.} \label{fig_gen_gdh_pio}
\end{figure}

As already noted, we used MAID2003 to calculate the elementary pion electroproduction
amplitude, which is sandwiched between the $NN$ initial
and final state wave functions. The usual calculational method, taking into account the
interaction between the final particles, follows the scheme
\begin{equation}\label{T}
T=T^{IA}+T^{NN}+T^{\pi N}\,,
\end{equation}
where $T^{IA}$ corresponds to the pure spectator model, whereas the other two terms
include $NN$ and $\pi N$ rescatterings treated up to the first order in the corresponding
two-body $t$-matrices $t_{NN}$ and $t_{\pi N}$.

For the deuteron wave function as well as for the final two nucleon state we used the
separable representation of the Paris potential from Ref.~\cite{Haid}. This model fits
the $NN$-phases up to a lab kinetic energy of 330~MeV. Although in the final $NN$
subsystem also energies above this value appear in the calculation, the contribution of
such events is insignificant as has been shown in~\cite{ArF05}, so that the
implementation of the model~\cite{Haid} is justified. For the $\pi N$ scattering matrix
we also used the separable representation from Ref.~\cite{NBL}.

\begin{figure}[htbp]
\centerline{\includegraphics[width=.49\textwidth]{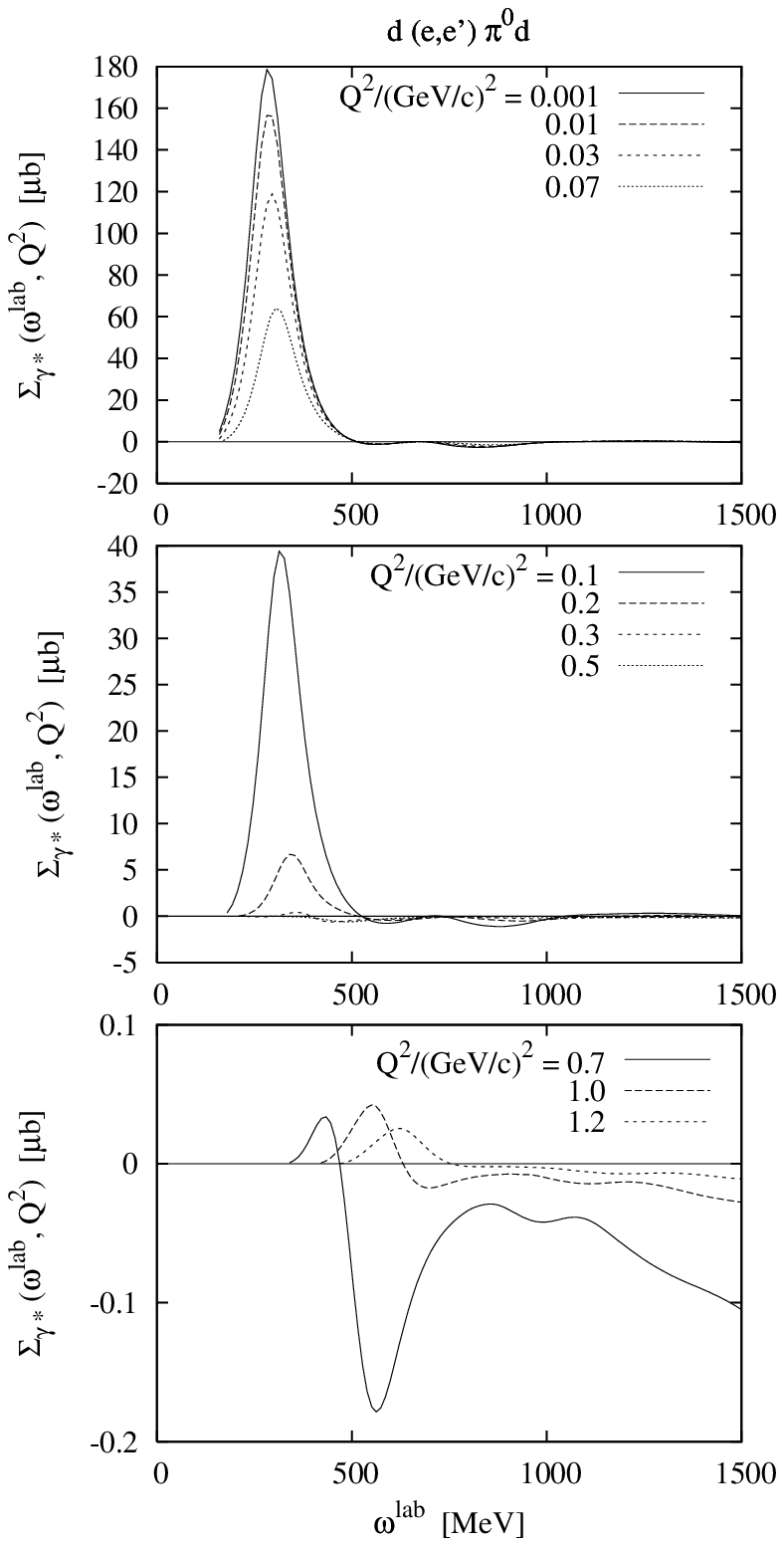}
\includegraphics[width=.49\textwidth]{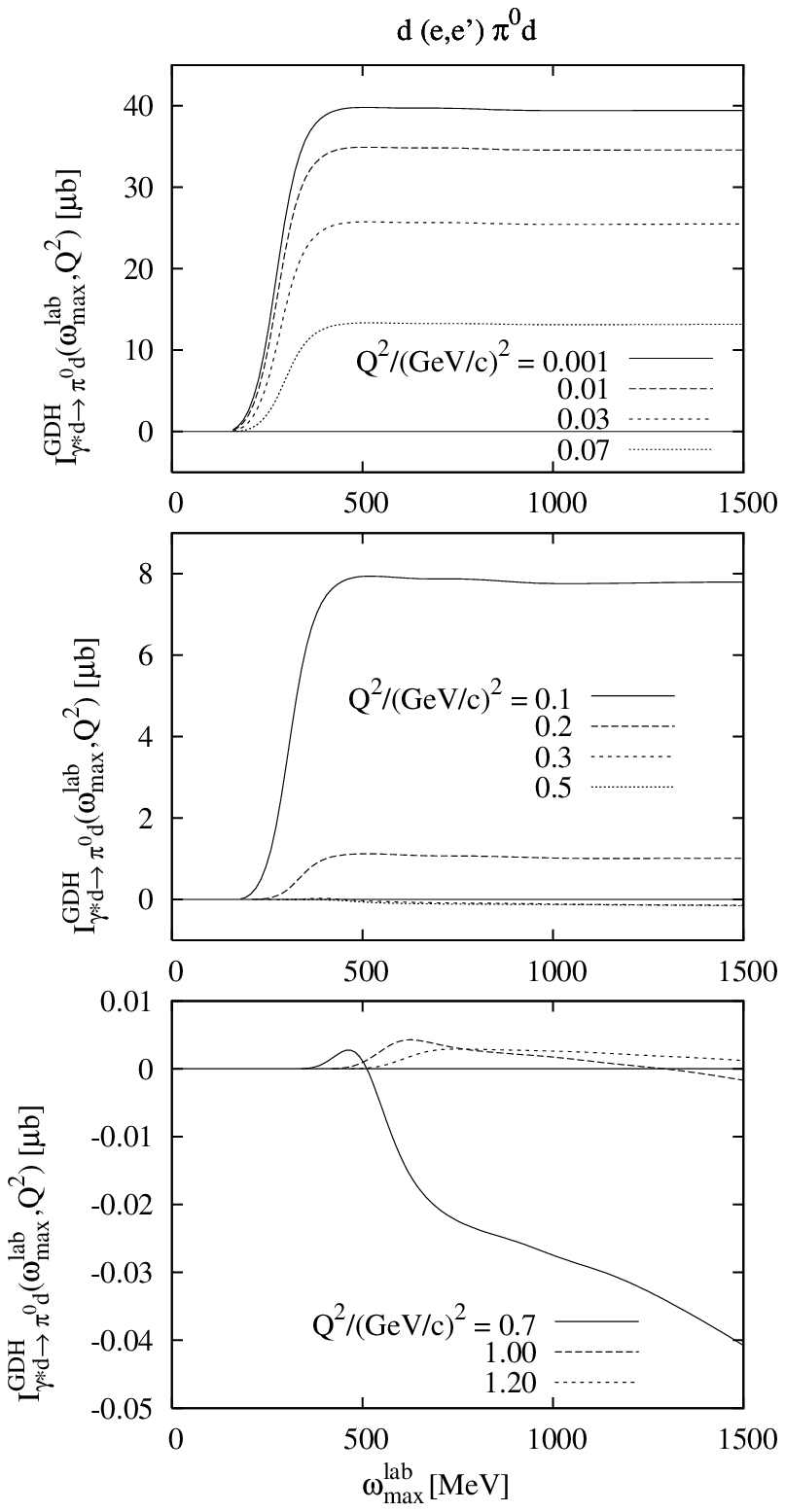}}
\caption{Transverse spin asymmetry $\Sigma_{\gamma^*d\to\pi^0d }$  as
  function of $\omega^{{lab}}$ (left panel)
  and finite GDH integral $I^{GDH}_{\gamma^*d\to\pi^0d}$ as function of
  $\omega^{{lab}}_{max}$ (right panel) of coherent $\pi^0$
  electroproduction on the deuteron
$d(e,e'\pi^0)d$ for various constant squared
four-momentum transfers $Q^2$.} \label{fig_gen_gdh_piocoh}
\end{figure}

\section{Results for spin asymmetry and finite GDH integral}
\label{sec4a}

\begin{figure}[htbp]
\centerline{\includegraphics[width=.49\textwidth]{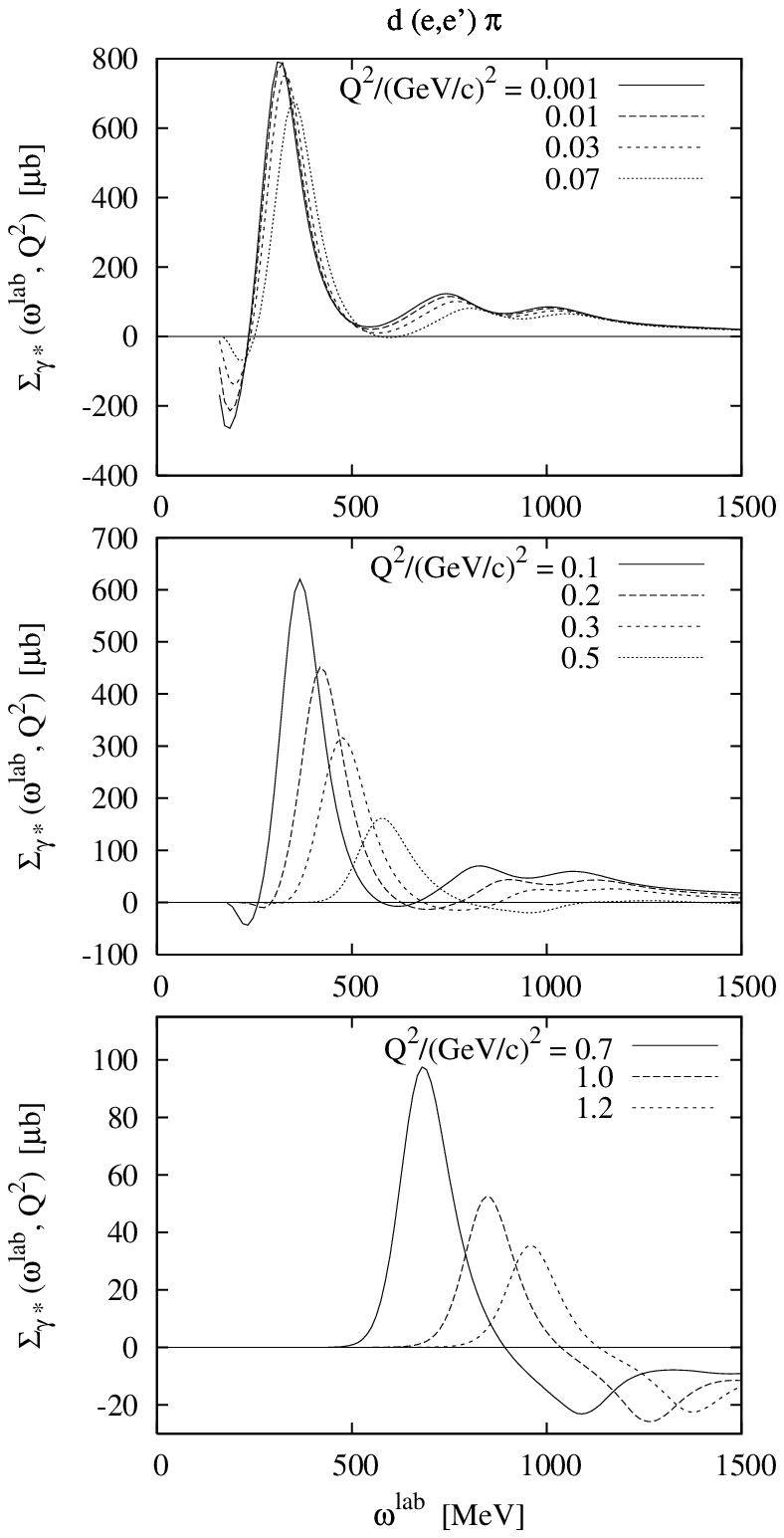}
\includegraphics[width=.49\textwidth]{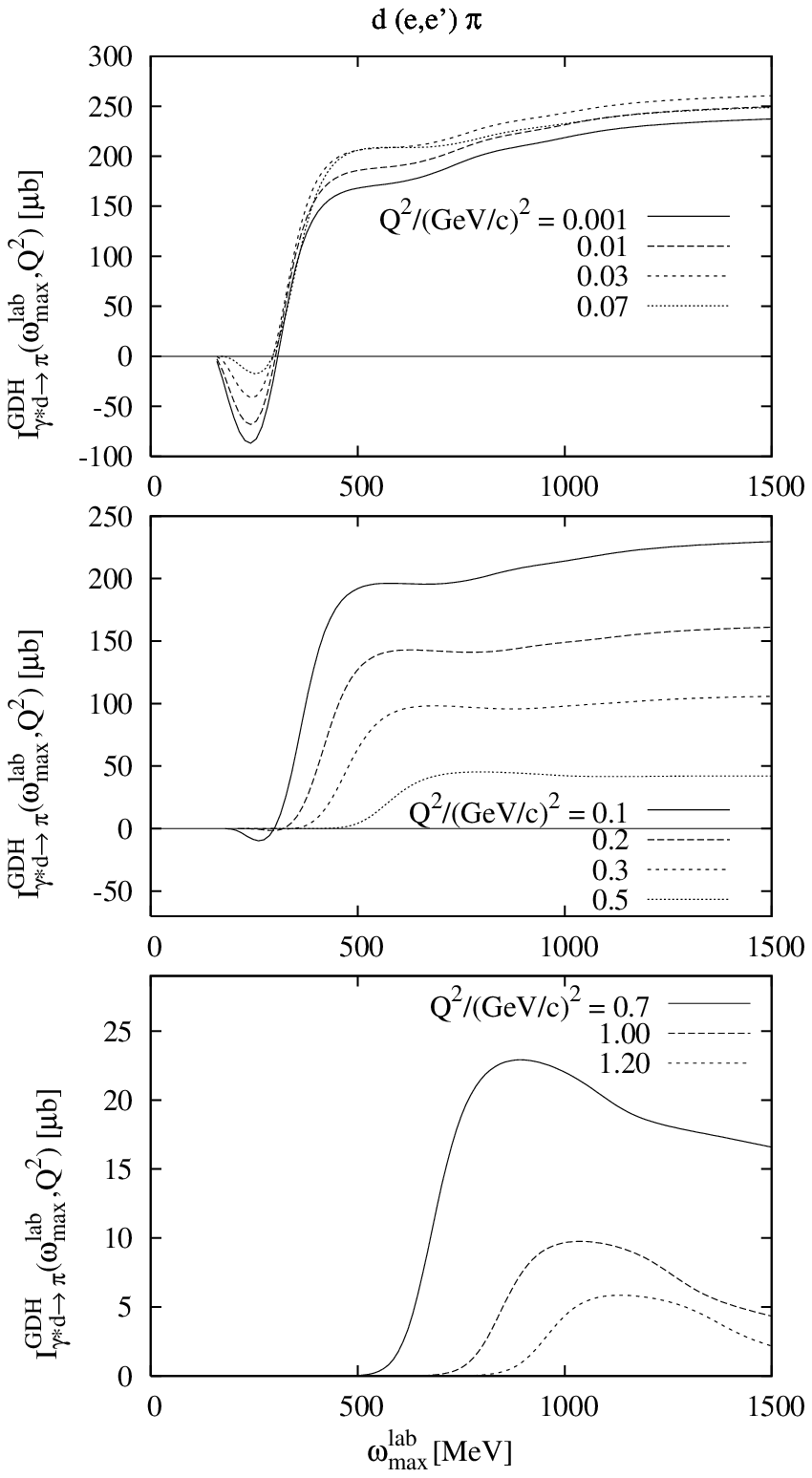}}
\caption{Total transverse spin asymmetry $\Sigma_{\gamma^*d\to\pi}$ as
  function of $\omega^{{lab}}$ (left panel) and finite GDH
integral $I^{GDH}_{\gamma^*d\to\pi}$ as function of
  $\omega^{{lab}}_{max}$ (right panel) of single pion
  electroproduction on the deuteron $d(e,e'\pi)$ for various constant
  squared four-momentum transfers
$Q^2$.} \label{fig_gen_gdh_piall}
\end{figure}

\begin{figure}[htbp]
\centerline{\includegraphics[width=.99\textwidth]{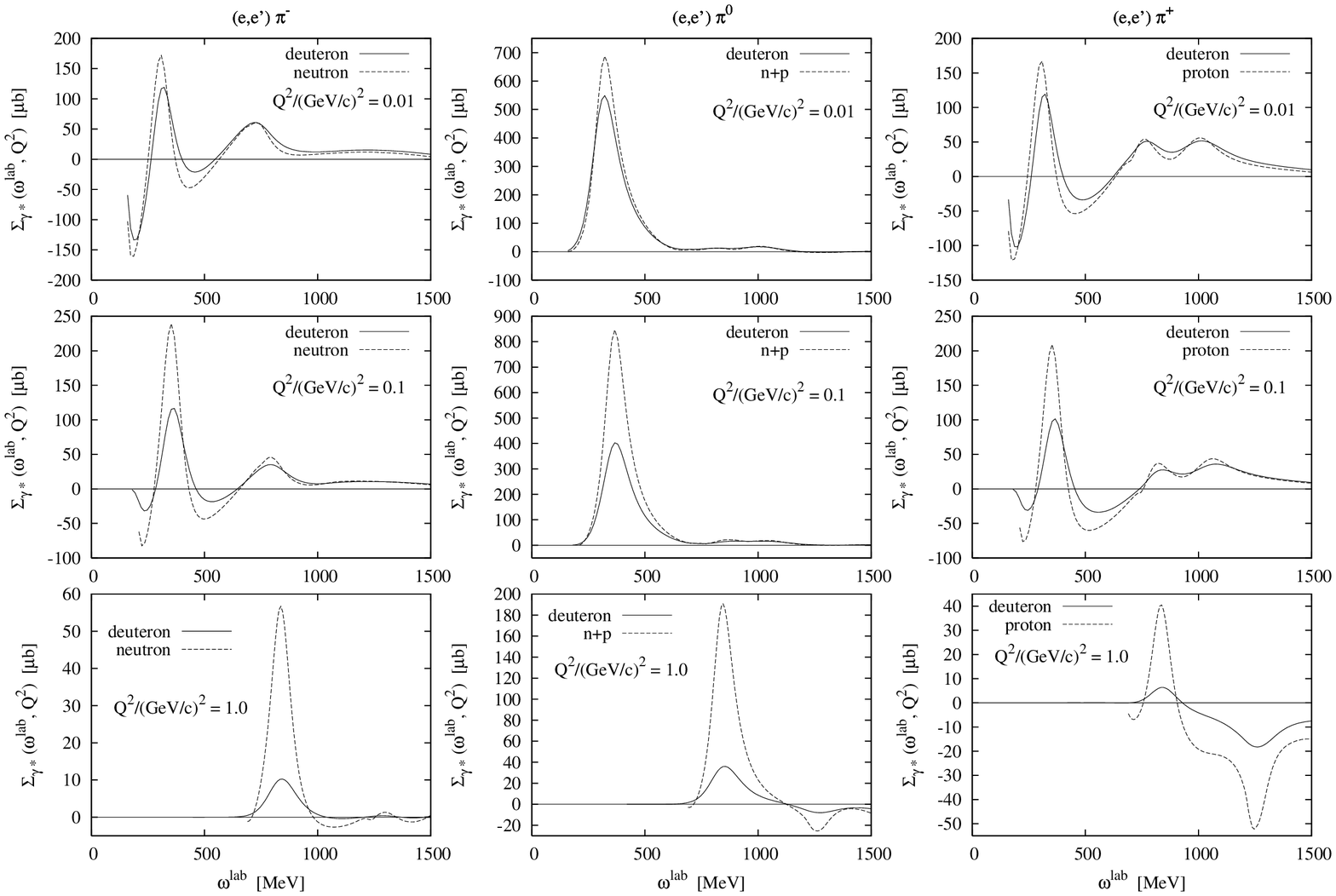}}
\caption{Total transverse spin asymmetries $\Sigma_{\gamma^*}$ for single $\pi^-$ (left
panel), $\pi^0$ (middle panel), and $\pi^+$ (right panel) electroproduction on deuteron
and nucleon as function of $\omega^{{lab}}$ for various constant squared
four-momentum transfers $Q^2$. For $\pi^0$ the solid curve for deuteron comprises the sum
of coherent and incoherent production.}
\label{fig_gen_gdh_piall_comparison}
\end{figure}

In the left panels of Figs.~\ref{fig_gen_gdh_pim} through \ref{fig_gen_gdh_piocoh} we
present our results for the spin asymmetry $\Sigma_{\gamma^*}(\omega^{lab},Q^2)$ as
defined in~(\ref{spin_asy_virtual}) for charged and neutral pion production as function
of the photon energy $\omega^{lab}$ for different values of the squared four-momentum
transfer $Q^2$. The right panels of these figures exhibit the finite GDH-integral as
function of the upper integration limit $\omega^{lab}_{max}$
in order to check the convergence of $I_{\gamma^*
d\to\pi}^{GDH}(Q^2,\omega^{lab}_{max})$ within the restricted energy domain of the present
work. The corresponding results on the asymmetry and the finite GDH-integral for the sum
of all channels are exhibited in Fig.~\ref{fig_gen_gdh_piall}. Furthermore, we show in
Fig.~\ref{fig_gen_gdh_piall_comparison} a comparison of the transverse spin asymmetries
of single pion production on nucleon and deuteron for $Q^2=0.01$, 0.1 and
1.0~(GeV/c)$^2$.

Before turning to the discussion we note that the impulse approximation $T^{IA}$
(first term in eq.~(\ref{T})) provides quite an adequate description of pion
photoproduction in the incoherent channels, whereas two-body mechanisms are of little
importance, at least not very close to the threshold. At the same time, it is reasonable
to expect that the amplitude $T^{IA}$ has only a weak sensitivity to the details of the
deuteron wave function as long as $Q^2$ is not too large. Therefore,
the total spin asymmetry in the incoherent channel should obey
approximately the simple relation
\begin{equation}\label{11a}
\sigma^P_{T,\gamma^*}-\sigma^A_{T,\gamma^*}\approx
(\sigma^p_{3/2}+\sigma^n_{3/2})-(\sigma^p_{1/2}+\sigma^n_{1/2})\,,
\end{equation}
which is in line with the assumption that the nucleon spins are aligned along the
deuteron spin except for a small pollution by the presence of the $D$-state. The relation
(\ref{11a}) rests furthermore on the assumption that Fermi motion, FSI and other two-body
effects can be disregarded and furthermore interference effects between proton and neutron
amplitudes can be neglected. However, already at the photon point as
well as for low $Q^2$ values one notes significant deviations (see
Ref.~\cite{AreFiSch} and Fig.~\ref{fig_gen_gdh_piall_comparison}) for
which the Fermi motion is mainly responsible. Moreover, in the low
energy region and at higher values of $Q^2$ one may expect that short
range mechanisms, where two-body effects are expected to become
important, play a more and more significant role. Therefore, in
these regions larger deviations from (\ref{11a}) may occur.

From Figs.~\ref{fig_gen_gdh_pim} through \ref{fig_gen_gdh_piocoh} one readily notes, that
in the region of low $Q^2$ the general energy dependence of the asymmetry becomes similar
to that of photoproduction~\cite{AreFiSch}. Just above the threshold, due to the large
wave length of the virtual photon the main mechanism of the reaction is described in
terms of dipole transitions, i.e.\ $E1$ and $M1$.
At low energies the $E1$ amplitude leads to $s$-wave pion production primarily via
the Kroll-Ruderman term with a small $d$-wave correction. As a consequence, $\pi^+$ and
$\pi^-$ production (Figs.~\ref{fig_gen_gdh_pim} and \ref{fig_gen_gdh_pip}) is mainly
governed by a strong $\sigma^A_{T,\gamma^*}$ contribution dominating over
$\sigma^P_{T,\gamma^*}$ up to energies of about $\omega^{lab}=200$ MeV. In this region
the values of $\Sigma_{\gamma^*}$ for the charged channels $\pi^-pp$ and $\pi^+nn$ are
comparable in magnitude. The small difference is due to the different dipole moments of
the $\pi^+n$ and $\pi^-p$ systems (because of the vanishing charge of the neutron), so
that the approximate relation holds
\begin{equation}
\frac{\Sigma_{\gamma^*d\to\pi^+nn}^{E1}}{\Sigma_{\gamma^*d\to\pi^-pp}^{E1}}
\approx\frac{1}{(1+m_\pi/M_N)^2}\approx 0.77\,,
\end{equation}
in agreement with the results shown in Figs.~\ref{fig_gen_gdh_pim} and
\ref{fig_gen_gdh_pip}.

In the neutral channels (Figs.~\ref{fig_gen_gdh_pio} and \ref{fig_gen_gdh_piocoh}), due
to the small pion mass resulting in a vanishingly small dipole moment of $\pi^0 N$, the
$E1$ transition is strongly suppressed below the first resonance and the corresponding
spin asymmetries $\Sigma_{\gamma^*d\to\pi^0np}$ and $\Sigma_{\gamma^*d\to\pi^0d}$ are
comparable with zero. This property is seen at all values of $Q^2$.

In the first resonance region the spin structure is mainly governed by the incoherent sum
of $M1$ and $E1$ transitions originating from the $P_{33}(1232)$ electroexcitation and
the Born terms, respectively. In the neutral channel $\pi^0np$ the electric transitions
remain insignificant up to an energy $\omega^{lab}=700$~MeV, where the $D_{13}(1520)$
resonance is exited via absorption of an $E1$ photon. An additional very small
contribution comes from the electric $E1$ component of the convection current due to the
Fermi motion of the bound nucleons. As a result, in the whole energy region up to
$\omega^{lab}=500$ MeV the electroproduction of $\pi^0$ proceeds almost exclusively via
the $M1$ transition to the $P_{33}(1232)$ resonance, which is especially well seen in
this channel. This mechanism is slightly enhanced by the nucleon pole terms in the $s$-
and $u$-channels, coming from the magnetic coupling of the photon to the nucleon.

\begin{figure}[htbp]
\centerline{\includegraphics[width=.89\textwidth]{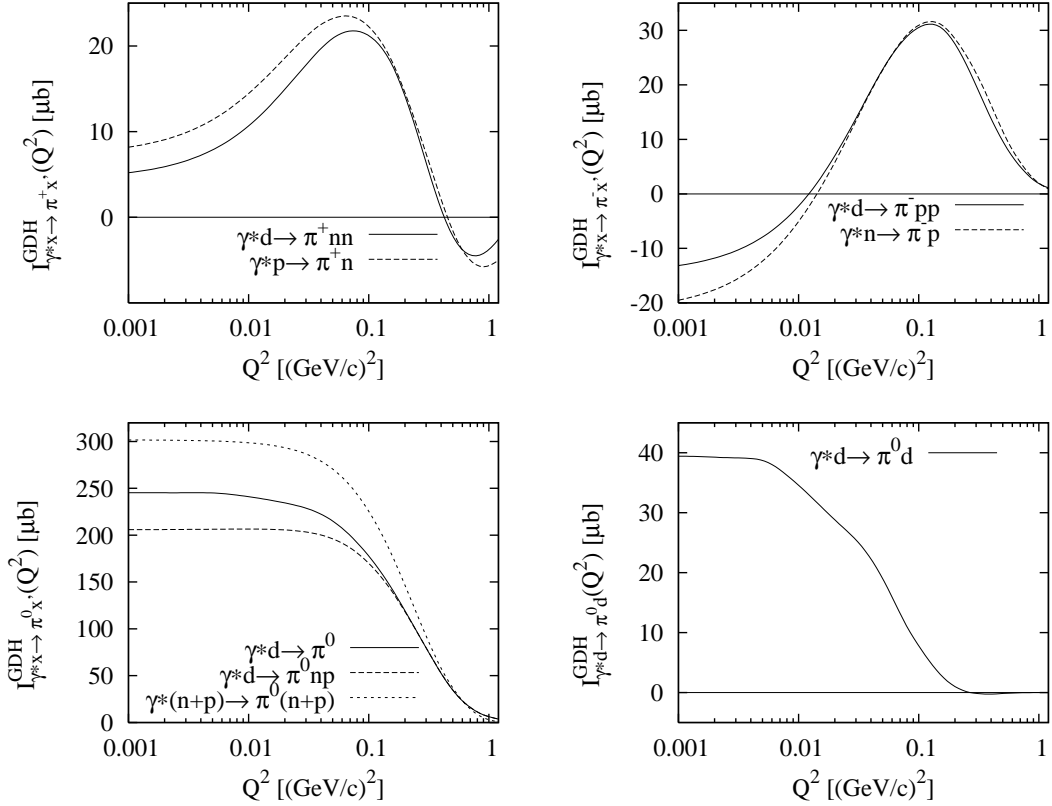}}
\caption{Finite GDH integrals for the separate channels of single pion
  electroproduction on the
deuteron $d(e,e'\pi)$ integrated up to $\omega^{{lab}_{max}}=1.5$~GeV as function of
squared four-momentum transfer $Q^2$.} \label{fig_gen_gdh_int_piall}
\end{figure}

Starting from $\omega^{lab}=600$~MeV, the behavior of $\Sigma_{\gamma^*}$ is mainly
determined by the localization of the dipole $E1$ strength in the region of the
$D_{13}(1520)$ and the quadrupole $E2$ transition via excitation of $F_{15}(1680)$. As a
result, in $\pi^+nn$ and $\pi^0np$ the curves exhibit two peaks at $\omega^{lab}=780$~MeV
and 1~GeV. The latter is almost invisible in the $\pi^-pp$ channel due
to a relatively weak coupling of the $F_{15}(1680)$ to $\gamma
n$. Above the second resonance region the spin
asymmetry remains small in all channels and demonstrates quite a smooth behavior.

The $Q^2$ evolution of $\Sigma_{\gamma^*}$ is partially determined by the $Q^2$
dependence of the spin asymmetry of the elementary nucleon reactions, which in our case
is given by the MAID2003 parametrization used in the present work. In particular,
nucleon Born terms and vector meson exchange are parametrized with a standard dipole form
factor. It is worth noting, that above the $P_{33}(1232)$ resonance
the strong background in the charged channels is to a
large extent canceled in the spin asymmetry $\Sigma_{\gamma^*}$.

\begin{figure}[htbp]
\centerline{\includegraphics[width=.7\textwidth]{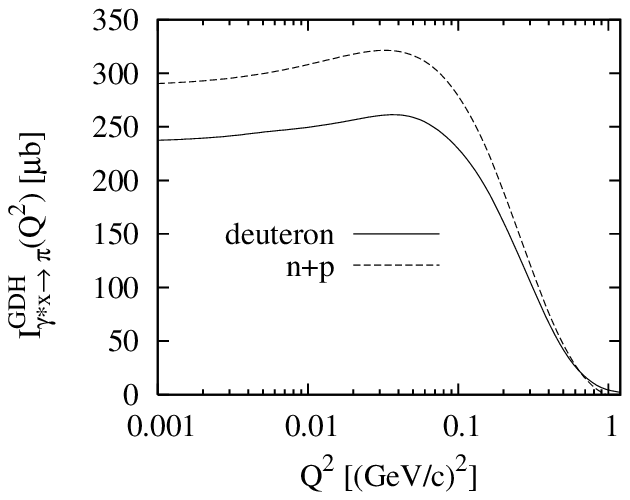} }
\caption{Total finite GDH integral of single pion electroproduction on
  the deuteron $d(e,e'\pi)$ (solid curve) and on neutron plus proton (dashed curve)
integrated up to $\omega^{lab}_{max}=1.5$~GeV as function of squared four-momentum
transfer $Q^2$.} \label{fig_gen_gdh_int_Q2}
\end{figure}

As for the resonance sector, the corresponding experimental information on the $Q^2$
dependence is still quite scarce, even for the transverse helicity components $A_\lambda$
with $\lambda=1/2,\,3/2$. In the region of low $Q^2$ one may hope that the largest
contribution to the sum rule is still provided by the low lying nucleon resonances, in
particular by the $P_{33}(1232)$ resonance whose internal spatial structure is quite well
understood. This is however not the case for higher values of $Q^2$ where higher
resonances start to come into play. Therefore the MAID2003 analyses still leaves room for
some variation of $\Sigma_{\gamma^*}(\omega^{lab},Q^2)$.

\begin{figure}[htbp]
\centerline{\includegraphics[width=.6\textwidth]{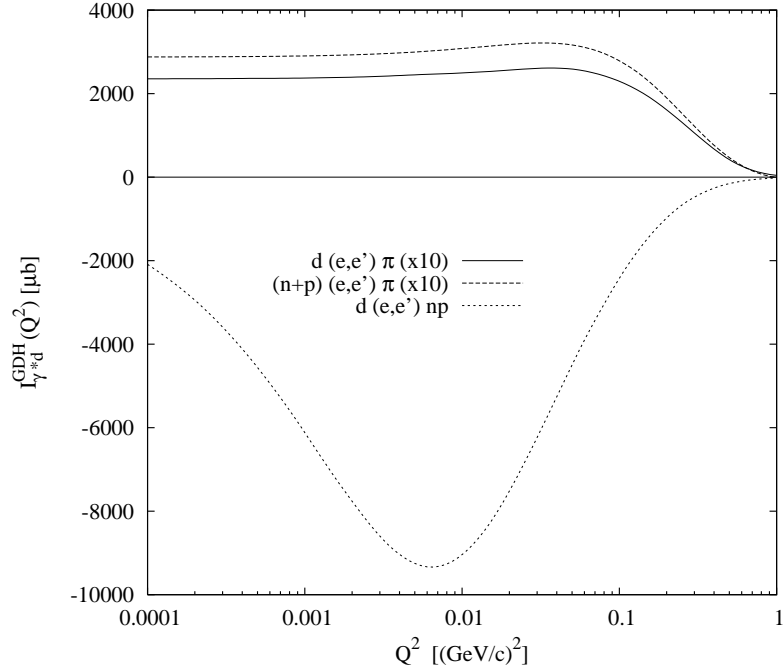}}
\caption{Total GDH integral of single pion electroproduction on the deuteron $d(e,e'\pi)$
(solid curve), electrodisintegration $d(e,e')np$ (dotted curve), and
on neutron plus proton (dashed curve) as function of squared
four-momentum transfer $Q^2$. The curves representing pion production
(solid and dashed curves) are magnified by a factor 10.}
\label{fig_gen_gdh_int_v18_piall}
\end{figure}

The $Q^2$ dependence of the $N\to P_{33}(1232)$ transition was studied in a wide range
and is shown to have an
almost constant ratio of the electric to the magnetic components. As a result, the
contribution of $E2$ remains vanishingly small and the relation
\begin{equation}\label{12a}
\frac{\sigma^{P(M1)}_{T,\gamma^*}}{\sigma^{A(M1)}_{T,\gamma^*}}\approx
\frac{\sigma^{3/2}}{\sigma^{1/2}}\approx 3\,.
\end{equation}
holds over a wide region of $Q^2$.

Quite remarkable is the rapid decrease of $\Sigma_{\gamma^*}(Q^2)$ in the second and the
third resonance region. This is especially well seen in the $\pi^+nn$ channel where,
starting from about $Q^2=0.3$ (GeV/c)$^2$, the asymmetry becomes predominantly negative.
This increasing relative contribution of the antiparallel component
$\sigma^A_{T,\gamma^*}$ especially
seen in $\pi^+$ photoproduction was also noted in Ref.~\cite{DrT04}. The helicity
amplitudes $A_\lambda$ ($\lambda=1/2,\,3/2$) of MAID give for the transitions $N\to
D_{13}(1520)$ and and $N\to F_{15}(1680)$ respectively
\begin{equation}\label{13a}
\left(\frac{A^{p(D_{13})}_{3/2}}{A^{p(D_{13})}_{1/2}}\right)^2\approx
31\, (48\pm 36)\,,\quad
\left(\frac{A^{p(F_{15})}_{3/2}}{A^{p(F_{15})}_{1/2}}\right)^2\approx 29\, (79\pm 64)\,,
\end{equation}
where the corresponding PDG values from~\cite{PDG} are given in parentheses. As a result,
at low $Q^2$ we see a strong dominance of the parallel component $\sigma^P_{T,\gamma^*}$.
As is discussed in~\cite{DrT04}, with
increasing virtuality of the photon, the amplitude $A_{3/2}$ drops faster than $A_{1/2}$, so that the resulting helicity asymmetry eventually runs through a
zero. In the $\pi^+nn$ channel this effect even forces the integral
$I^{GDH}_{\gamma^*d\to\pi^+nn}(Q^2)$ to become negative at about $Q^2=0.3$~(GeV/c)$^2$
(upper left panel in Fig.~\ref{fig_gen_gdh_int_piall}). Another reason for the
predominance of $\sigma^A_{T,\gamma^*}$ at higher $Q^2$ should be a rather smooth $Q^2$
dependence of the $S_{11}(1535)$, so that this resonance, contributing exclusively to the
antiparallel component, remains excited at rather large values of momentum transfer.

Finally, in the region $Q^2>0.5$~(GeV/c)$^2$, in the absence of the main contribution
from the first and partially the second resonance region, higher resonance states start
to dominate the GDH integral. As a result, its value becomes sensitive to the behavior of
the cross section at higher energies. An important consequence of this fact is, as
already noted above, that the integral $I_{\gamma^*d\to\pi
  NN}^{GDH}(\omega^{lab}_{max},Q^2)$ has not reached
convergence for these higher $Q^2$-values within the energy region of
our study. Therefore, one needs in future studies to extend the
elementary multipole analysis at least up to lab energies above 2 GeV. For the neutral
channels in comparison to $\pi^{\pm}$ the convergence appears to be better for the low
$Q^2$-values whereas again one notes at higher $Q^2$ that convergence has not been
reached.

The resulting values of the GDH integral as function of $Q^2$ in different channels are
presented in Fig.~\ref{fig_gen_gdh_int_Q2}. As is noted in~\cite{Korchin}, the maximum in
$I_{\gamma^*d\to\pi NN}^{GDH}(Q^2)$ for $\pi^+nn$ and $\pi^-pp$ seen at about
0.05~(GeV/c)$^2$ is simply due to a different $Q^2$ dependence of the Kroll-Ruderman and
the $P_{33}(1232)$ terms. Namely, the $P_{33}(1232)$ contribution
dominating the parallel component
$\sigma^P_{T,\gamma^*}$ at lower $Q^2$ changes quite slowly up to a value where
$\sqrt{Q^2}$ becomes comparable to the $\rho$ meson mass (see the experimental results
compiled in~\cite{MaidR}). Above this point, the $P_{33}(1232)$ form factor visibly
suppresses the $P_{33}(1232)$ resonance peak. On the contrary, the $\sigma^A_{T,\gamma^*}$
component, dominated by the Kroll-Ruderman term, behaves like the inverse momentum of the
incident virtual photon, and thus rapidly vanishes. This interplay leads to the slow
increase of $I_{\gamma^*d\to\pi}^{GDH}(Q^2)$ up to about $Q^2=0.05$~(GeV/c)$^2$ with a
subsequent fall-off forced by the strong $Q^2$ dependence of the
$P_{33}(1232)$ form factor.

Coherent pion photoproduction on the deuteron in the first resonance region is almost
totally determined by the spin independent part of the $M1$ transition to the
$P_{33}(1232)$ resonance. As a result, the cross section is strongly dominated by the
$\sigma^P_{T,\gamma^*}$ component. The resonance $P_{33}(1232)$ is especially well seen
in this channel up to quite high $Q^2$ values, and the GDH integral
(\ref{fig_gen_gdh_int_piall}) is saturated at already rather low $\omega^{lab}$. At the
same time, the extended structure of the deuteron results in quite a rapid fall-off of
$I_{\gamma^*d\to\pi^0d}^{GDH}(Q^2)$ with increasing $Q^2$ via the deuteron form factor.

Finally, we show in Fig.~\ref{fig_gen_gdh_int_v18_piall} a comparison of the GDH integral
of single pion production with the one of electrodisintegration. One
readily notes for the disintegration channel the
pronounced deep negative minimum near $Q^2=0.006$~(GeV/c)$^{2}$,
which we had mentioned in the introduction, whereas the pion production channel exhibits
below $Q^2=0.1$~(GeV/c)$^{2}$ a nearly constant contribution of,
however, much smaller size,
and then a rapid fall-of at higher $Q^2$ values. Thus the disintegration channel remains
the dominant feature for the generalized GDH sum rule.

\section{Summary and conclusions}
\label{sec4}

The beam-target spin asymmetry of single pion electroproduction on the
deuteron for transverse virtual photons and the associated generalized
Gerasimov-Drell-Hearn integral have been evaluated by explicit
integration up to an energy $\omega^{{lab}}=1.5$~GeV and for
squared momentum transfers between 0.001 and 1.2~(GeV/c)$^2$.
Whereas below pion production threshold the main contribution (negative) to the
transverse spin asymmetry from electrodisintegration comes from the
strong $M1$ transition to the resonant $^1S_0$ scattering state near the
disintegration threshold (so-called antibound state), mainly
driven by the large nucleon anomalous magnetic moment with additional meson exchange
current and relativistic contributions~\cite{Are04}, above the threshold single nucleon
mechanisms, primarily single pion electroproduction start to dominate the spin asymmetry.

According to our results, the contribution of pion production to
$I_{\gamma^*d\to \pi}^{GDH}(Q^2)$ coming from
the energy region below $\omega^{lab}=1.5$ GeV is to a large extent saturated by
resonance electroexcitations. The calculation based on the MAID2003 model for the
elementary pion production amplitude shows that at low $Q^2$ the dominant contribution
comes from the $P_{33}(1232)$ resonance. As the virtuality of the photon increases, the
role of higher resonances tends to be more and more important.
This effect is in addition amplified by the damping of the
$P_{33}(1232)$ at higher $Q^2$ as well as by a rapid increase of the relative
contribution of the antiparallel component $\sigma^A_{T,\gamma^*}$ above the first
resonance. As a result, we find a strong $Q^2$ dependence of the integral
$I^{GDH}_{\gamma^*d\to\pi}$, so that at $Q^2=1$~(GeV/c)$^2$ it comprises only about
5~$\mu b$.

In general, our calculation shows that for $Q^2\leq 0.5$~(GeV/c)$^2$ the major
contribution of single pion electroproduction to $I^{GDH}_d(Q^2)$ is
contained in the region $\omega^{lab}\leq 1.5$~GeV. For higher values of $Q^2$ the
finite GDH integral does not exhibit good convergence in the
considered energy range pointing to the need for an extension of the
present analysis beyond the energy region considered in this work. Moreover,
very likely one would also need to consider two pion production.

As for the question about the behavior of the total sum rule
$I_{\gamma^*d}^{GDH}(Q^2)$ comprising disintegration and pion production
at finite $Q^2$, we see that with increasing $Q^2$ the single pion
photoproduction does not compensate the rapid
change of the negative contribution of the nucleonic channel
$I^{GDH}_{\gamma^*d\to np}(Q^2)$, so
that the resulting total integral $I_{d}^{GDH}(Q^2)$ exhibits
almost the same strong $Q^2$ dependence as the disintegration channel alone. It is
very unlikely that multiple meson production, which is so far not
included in this study, will be able to visibly change the value of
$I_{d}^{GDH}(Q^2)$. Therefore, the strong dominance
of the disintegration channel, leading to a large negative value of
$I_{d}^{GDH}(Q^2)$ below about $Q^2=1$~(GeV/c)$^{2}$ seems to
remain a special feature of the generalized GDH sum rule
of the deuteron.


\begin{thebibliography}{99}

\bibitem{Ger65} S.~B.~Gerasimov, Yad.\ Fiz.\ {\bf 2}, 598 (1965)
                (Sov.\ J.\ Nucl.\ Phys.\ {\bf 2}, 430 (1966)).

\bibitem{DrH66} S.~D.~Drell and A.~C.~Hearn, Phys.\ Rev.\ Lett.\
                {\bf 16}, 908 (1966).

\bibitem{OJahn}
   J.~Ahrens {\it et al.}, Phys.\ Rev.\ Lett.\ {\bf 97}, 202303 (2006).

\bibitem{Dre01} D.~Drechsel,
    Proc. Symposium on the GDH Sum Rule,
    Mainz 2000, eds. D. Drechsel and L. Tiator (World Scientific,
    Singapore 2001).

\bibitem{DrT04} D.~Drechsel and L.~Tiator, Ann.\ Rev.\ Nucl.\ Sci.\ {\bf 54}, 69
  (2004).

\bibitem{AreFiSch}
  H.~Arenh\"ovel, A.~Fix and M.~Schwamb, Phys.\ Rev.\ Lett.\  {\bf 93}, 202301 (2004).

\bibitem{FiA2pi}
  A. Fix and H. Arenh\"ovel, Eur.\ Phys.\ J.\  A {\bf 25}, 115 (2005).

\bibitem{Anselm}
  M.~Anselmino, B.~L.~Ioffe, and E.~Leader,
  Sov.\ J.\ Nucl.\ Phys.\  {\bf 49}, 136 (1989).

\bibitem{Burkert:1992yk}
  V.~Burkert and Z.~J.~Li,
  Phys.\ Rev.\  D {\bf 47}, 46 (1993).

\bibitem{Are04}
  H.~Arenh\"ovel, Phys.\ Lett.\ B {\bf 595}, 223 (2004).

\bibitem{TaF06}
  M.~Tammam, A.~Fix, and H.~Arenh\"ovel, Phys.\ Rev.\ C {\bf 74}, 044001
  (2006).

\bibitem{MAID03}
        D. Drechsel, O. Hanstein, S.S. Kamalov, and L. Tiator,
        MAID: http://www.kph.uni-mainz.de/de/MAID/.

\bibitem{ArF05}
  H.~Arenh\"ovel and A.~Fix, Phys.\ Rev.\ C {\bf 72}, 064004 (2005).

\bibitem{Haid}
   J.~Haidenbauer and W.~Plessas, Phys.\ Rev.\ C {\bf 30}, 1822
   (1984); {\bf 32}, 1424 (1985).

\bibitem{NBL}
   S.~Nozawa, B.~Blankleider, and T.-S.~H.~Lee, Nucl.\ Phys.\ A {\bf 513}, 459 (1990).

\bibitem{PDG}
  C.~Amsler {\it et al.} [Particle Data Group], Phys.\ Lett.\ B {\bf 667}, 1 (2008).

\bibitem{Korchin}
  O.~Scholten and A.~Y.~Korchin, Eur.\ Phys.\ J.\  A {\bf 6}, 211 (1999).

\bibitem{MaidR}
  L.~Tiator, D.~Drechsel, S.~Kamalov, M.~M.~Giannini, E.~Santopinto, and A.~Vassallo,
  Eur.\ Phys.\ J.\  A {\bf 19}, 55 (2004).

\end{thebibliography}
\end{document}